\documentclass[twocolumn,showpacs,preprintnumbers,amsmath,amssymb]{revtex4}

\usepackage{graphicx}
\usepackage{dcolumn}
\usepackage{bm}

\begin{document}

\preprint{}
\input{epsf.tex}

\epsfverbosetrue

\title{Collective Electronic Excitation Coupling between Planar Optical Lattices using Ewald's Method}

\author{Hashem Zoubi, and Helmut Ritsch}

\affiliation{Institut fur Theoretische Physik, Universitat Innsbruck, Technikerstrasse 25, A-6020 Innsbruck, Austria}

\date{06 June, 2011}

\begin{abstract}
Using Ewald's summation method we investigate collective electronic excitations (excitons) of ultracold atoms in parallel planar optical lattices including long range interactions. The exciton dispersion relation can then be suitably rewritten and efficiently calculated for long range resonance dipole-dipole interactions. Such in-plane excitons resonantly couple for two identical optical lattices, with an energy transfer strength decreasing exponentially with the distance between the lattices. This allows a restriction of the transfer to neighboring planes and gives rise to excitons delocalized between the lattices. In general equivalent results will hold for any planar system containing lattice layers of optically active and dipolar materials.
\end{abstract}

\pacs{37.10.Jk, 42.50.-p, 71.35.-y}

\maketitle

\section{Introduction}

Ultracold atoms in optical lattices can be considered as one of the most attractive quantum systems that occupied a wide range of theoretical and experimental researches \cite{Dalibard}. The controlability of the different system parameters makes it a test set-up for many condensed matter effects and models, and opened the door toward answering fundamental questions and getting a deep understanding of condensed matter phenomena \cite{Lewenstein}. Most recent research is restricted to atoms in their lowest hyperfine manifold. As a key phenomenon a quantum phase transition from a superfluid to a Mott insulator phase can be achieved by controlling the depth of the optical lattice potential in changing the laser intensity \cite{Jaksch,Greiner}.

Electronic excitations for ultracold atoms in optical lattices, which posses an energy many orders of magnitude bigger than the kinetic temperature of the atoms, are generally avoided in this context. Due to their short life time and the momentum recoil in emission or absorption processes of a photon, this energy can heat the system and destroy the stability of the Mott insulator phase. This view is limited by considering electronic excitations as independent entities that localized at each optical lattice site, which is the case for time-of-flight observation techniques \cite{Spielman}, and also for metrology in using clock states \cite{Katori}. These considerations, which rest on treating the optical lattice only as a tool to localized the independent atoms at a fixed zones in space, missed the novelty of the formation of collective electronic effects that can be induced through the direct electrostatic interactions among atoms at different sites and in exploiting the lattice symmetry.

In previous work we generalized optical lattice dynamics to include collective electronic excitations of the atoms \cite{ZoubiA,ZoubiB,ZoubiC}. For sufficient density such an excitation gets delocalized in the lattice due to resonance dipole-dipole interactions, where energy transfer among atoms at different lattice sites and in momentum space is represented by a wave that propagates in the lattice with a fixed wave vector. It can be considered as a quasi-particle termed exciton \cite{Davydov,Agranovich}. The exciton quasi-momentum is distributed over the whole lattice atoms so that at the emission or absorption of a photon involving an exciton the whole lattice recoils and not a single atom. Therefore excitons do not necessarily destroy to the quantum phases and in particular the Mott phase as they exhibit a suppressed recoil effect. Interestingly the life time of excitons in low dimensional optical lattices as calculated recently \cite{ZoubiD} can be very short due to superradiance for some polarizations and wave vectors or metastable for others. This fact provides a strong tool that allows the controlability of excitation life times in optical lattices. 

In the present paper we investigate electronic excitations for ultracold atoms in planar optical lattices in an improved exciton picture. While in previous works we neglected the long range part of the dipole-dipole interaction in a nearest neighbor approximation \cite{ZoubiA,ZoubiB,ZoubiC}, here we explicitly calculate the corrections from long range interactions. In addition we account for the anisotropic nature of excitons for a fixed but nonorthogonal direction of the excitation transition dipole. The explicit calculation can be performed using the Ewald's method \cite{Born}. 

Once the eigenstates for a single plane are found the treatment can be extended to the case of two identical parallel planes. Excitons in the two planes will couple and lead to resonant energy transfer. Even though the resonance dipole-dipole interaction among two atoms in each plane has an inverse cube dependence on the distance, a summation over the whole plane leads to an exponentially decrease of the effective exciton-exciton coupling. At small distances one can form hybrid excitons delocalized between the two lattices planes. The results can be generalized to a finite number of identical planar optical lattices to describe a quasi 3D setup.

The paper is organized as follows: in section 2 we present hybrid excitons for a system of two parallel and identical planar optical lattices. The exciton dispersions are calculated in sections 3 in using the long range resonance dipole-dipole interaction and in applying Ewald's method. This section includes the calculation details of the exciton matrix elements for inter-lattices. The conclusions are given in section 4. The appendix includes the calculation of the intra-lattice exciton matrix elements. 

\section{Excitons in Planar Optical Lattices}

We consider two parallel identical planar optical lattices at distance $b$ and lattice constant $a$ filled with atoms in the Mott insulator phase with one atom per site as in figure (1). The atoms are two-level systems with on-site electronic transition energy $E_A$. An on-site electronic excitation couples to other sites via resonant dipole-dipole interaction, where an excitation at one site decays and other site gets excited. The energy transfer can be among sites in the same plane or among different planes. The electronic excitation Hamiltonian is given by
\begin{equation}
H=\sum_{{\bf n},\alpha}E_A\ B_{{\bf n}\alpha}^{\dagger}B_{{\bf n}\alpha}+\sum_{{\bf nm},\alpha\beta}J_{\bf nm}^{\alpha\beta}\ B_{{\bf n}\alpha}^{\dagger}B_{{\bf m}\beta}.
\end{equation}
The indexes $({\bf n,m})$ run over the position of the in-plane sites, and $(\alpha,\beta)$ over the two planes which are denoted by $(1,2)$. Here $B_{{\bf n}\alpha}^{\dagger},\ B_{{\bf n}\alpha}$ are the creation and annihilation operators of an electronic excitation at plane $\alpha$ and site ${\bf n}$. They are taken to be boson operators with the commutation relation $[B_{{\bf n}\alpha},B_{{\bf m}\beta}^{\dagger}]=\delta_{\bf nm}\delta_{\alpha\beta}$. Using boson operators for two-level atoms is an approximation that hold for a single excitation or for low density of excitations, where we neglect saturation effects in which two excitations trying to excite the same site. The parameter $J_{\bf nm}^{\alpha\beta}$ is for the resonance dipole-dipole interaction among an excitation at site $({\bf n},\alpha)$ and site $({\bf m},\beta)$.

Defined by the lattice symmetry, delocalized eigenstates can be represented as waves that propagate in the lattice with in-plane wave vector ${\bf k}$ and are denoted excitons. Hence the following transformation into the exciton basis formally diagonalizes the Hamiltonian
\begin{equation}
B_{{\bf n}\alpha}=\frac{1}{\sqrt{N}}\sum_{\bf k}B_{{\bf k}\alpha}e^{i{\bf k}\cdot{\bf n}},
\end{equation}
where $N$ is the number of in-plane sites. The Hamiltonian casts into
\begin{equation}
H=\sum_{{\bf k},\alpha}E_A\ B_{{\bf k}\alpha}^{\dagger}B_{{\bf k}\alpha}+\sum_{{\bf k},\alpha\beta}J^{\alpha\beta}({\bf k})\ B_{{\bf k}\alpha}^{\dagger}B_{{\bf k}\beta},
\end{equation}
where we defined the exciton dynamical matrix
\begin{equation}
J^{\alpha\beta}({\bf k})=\sum_{\bf R}J^{\alpha\beta}({\bf R})e^{i{\bf k}\cdot{\bf R}}.
\end{equation}
The interaction parameter is a function of the distance between the two sites, that is $J_{\bf nm}^{\alpha\beta}=J^{\alpha\beta}({\bf n-m})$, where we defined $J^{\alpha\beta}({\bf n-m})=J^{\alpha\beta}({\bf R})$, with ${\bf R=n-m}$.

As we have two identical lattices, the Hamiltonian can also easily be diagonalized in the lattice indexes in using the transformation
\begin{equation}
B_{{\bf k}\pm}=\frac{B_{{\bf k}1}\pm B_{{\bf k}2}}{\sqrt{2}},
\end{equation}
to get the full free exciton Hamiltonian
\begin{equation}
H=\sum_{{\bf k},\nu}E_{ex}^{\nu}({\bf k})\ B_{{\bf k}\nu}^{\dagger}B_{{\bf k}\nu},
\end{equation}
with $(\nu=\pm)$, and the exciton eigenenergies are
\begin{equation}
E_{ex}^{\pm}({\bf k})=E_A+J({\bf k})\pm J^{\prime}({\bf k}),
\end{equation}
where we defined inside the same lattice the coupling parameter
\begin{equation}
J({\bf k})=J^{11}({\bf k})=J^{22}({\bf k}),
\end{equation}
and between different lattices the coupling parameter
\begin{equation}
J^{\prime}({\bf k})=J^{12}({\bf k})=J^{21}({\bf k}).
\end{equation}
Our main target now is to calculate the exciton dynamical matrices $J({\bf k})$ and $J^{\prime}({\bf k})$.

\begin{figure}[h!]
\centerline{\epsfxsize=8cm \epsfbox{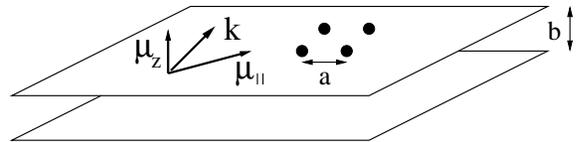}}
\caption{Two parallel planar optical lattices of lattice constant $a$, and separated by distance $b$, with one atom per site. The in-plane wave vector, the in-plane and normal transition dipole components are plotted.}
\end{figure}

\section{Exciton Dispersions}

Lets us now calculate the exciton energy dispersion explicitly for dipole-dipole interaction between two sites separated by a distance ${\bf R}$, which are given by
\begin{equation}
J({\bf R})=\frac{1}{4\pi\epsilon_0}\frac{|{\bf R}|^2|\mbox{\boldmath$\mu$}|^2-3(\mbox{\boldmath$\mu$}\cdot{\bf R})^2}{|{\bf R}|^5},
\end{equation}
where $\mbox{\boldmath$\mu$}$ is the electronic excitation transition dipole, with the components $\mbox{\boldmath$\mu$}=(\mu_x,\mu_y,\mu_z)$. The general distance between two sites is ${\bf R}=(l_xa,l_ya,l_zb)$, where the lattice is taken to be in the $(x-y)$ plane. We assume large square optical lattices with the number of in-plane sites $N$ and area $S=Na^2$. Hence, $l_x,l_y=0,\pm 1,\pm 2,\cdots,\pm\sqrt{N}/2$, with the limit of $N\gg 1$. As we consider here only two lattices, we have $l_z=0,1$, where one plane with $z=0$ and the other at $z=b$. The interaction parameter is written as
\begin{equation}
J({\bf R})=\frac{1}{4\pi\epsilon_0}\sum_{ij}\mu_i\mu_j\ D_{ij}({\bf R}),
\end{equation}
where $(i,j=x,y,z)$, with
\begin{equation} \label{DDI}
D_{ij}({\bf R})=\frac{\delta_{ij}}{\left(\bar{l}_x^2+\bar{l}_y^2+\bar{l}_z^2\right)^{3/2}}-3\ \frac{\bar{l}_i\bar{l}_j}{\left(\bar{l}_x^2+\bar{l}_y^2+\bar{l}_z^2\right)^{5/2}},
\end{equation}
where $\bar{l}_x=al_x$, $\bar{l}_y=al_y$ and $\bar{l}_z=bl_z$. In Fourier momentum space we get
\begin{equation}
J({\bf k})=\frac{1}{4\pi\epsilon_0}\sum_{ij}\mu_i\mu_j\ D_{ij}({\bf k}),
\end{equation}
where
\begin{equation}
D_{ij}({\bf k})=\sum_{\bf R}D_{ij}({\bf R})e^{i{\bf k}\cdot{\bf R}}.
\end{equation}
We have ${\bf k}=(k_x,k_y)$, which takes the values $k_{x,y}=\frac{2\pi}{a\sqrt{N}}p$, with $p=0,\pm 1,\pm 2,\cdots,\pm\sqrt{N}/2$.

From this point on we split the calculations into two parts, one for in-plane interactions, and the other for interactions among the two lattices. Here we concentrate in the exciton dynamical matrix elements for inter-lattices, while the intra-lattice case appears in the appendix.

For interactions between the two optical lattices we have $l_z=1$. We define the function
\begin{equation} \label{Oscil}
S({\bf k})=\sum_{l_x,l_y}\frac{e^{ia(k_xl_x+k_yl_y)}}{\left(a^2l_x^2+a^2l_y^2+b^2\right)^{5/2}}.
\end{equation}
Then we have the diagonal elements
\begin{eqnarray}
D_{xx}({\bf k})&=&\left(2\frac{\partial^2}{\partial k_x^2}-\frac{\partial^2}{\partial k_y^2}+b^2\right)S({\bf k}), \nonumber \\
D_{yy}({\bf k})&=&\left(2\frac{\partial^2}{\partial k_y^2}-\frac{\partial^2}{\partial k_x^2}+b^2\right)S({\bf k}), \nonumber \\
D_{zz}({\bf k})&=&\left(-\frac{\partial^2}{\partial k_x^2}-\frac{\partial^2}{\partial k_y^2}-2b^2\right)S({\bf k}),
\end{eqnarray}
and the off-diagonal elements
\begin{eqnarray}
D_{xy}({\bf k})&=&D_{yx}^{\ast}({\bf k})=3\frac{\partial^2}{\partial k_x\partial k_y}S({\bf k}), \nonumber \\
D_{xz}({\bf k})&=&D_{zx}^{\ast}({\bf k})=3ib\frac{\partial}{\partial k_x}S({\bf k}), \nonumber \\
D_{yz}({\bf k})&=&D_{zy}^{\ast}({\bf k})=3ib\frac{\partial}{\partial k_y}S({\bf k}).
\end{eqnarray}

We proceed in the calculation of the function $S({\bf k})$. Due to oscillations the series terms are slowly converging. Then we use Ewald's method to convert the oscillating series into a one with exponentially decaying terms \cite{Born}, and then we can concentrate in the first terms that represent small wave number or long wave length excitons.

In using the relation
\begin{equation} \label{Rel1}
\frac{4}{3\sqrt{\pi}}\int_0^{\infty}dt\ t^{3/2}e^{-ct}=\frac{1}{c^{5/2}},
\end{equation}
we obtain
\begin{eqnarray}
S({\bf k})&=&\frac{4}{3\sqrt{\pi}}\int_0^{\infty}dt\ t^{3/2}e^{-b^2t}\left\{\sum_{l_x}e^{iak_xl_x-a^2l_x^2t}\right\} \nonumber \\
&\times&\left\{\sum_{l_y}e^{iak_yl_y-a^2l_y^2t}\right\}.
\end{eqnarray}
We use the identity, for $(j=x,y)$,
\begin{equation} \label{Rel2}
\sum_{l_j}e^{iak_jl_j-a^2l_j^2t}=\frac{\sqrt{\pi}}{a\sqrt{t}}\sum_ne^{-\frac{1}{a^2t}\left(\pi n+\frac{ak_j}{2}\right)^2},
\end{equation}
to get
\begin{equation}
S({\bf k})=\frac{4\sqrt{\pi}}{3a^2}\int_0^{\infty}dt\ t^{1/2}e^{-b^2t}\sum_{n,m}e^{-\frac{\Gamma^2_{nm}}{a^2t}},
\end{equation}
where
\begin{equation}
\Gamma^2_{nm}=\left(\pi n+\frac{ak_x}{2}\right)^2+\left(\pi m+\frac{ak_y}{2}\right)^2.
\end{equation}
By applying the result
\begin{equation}
\int_0^{\infty}dt\ \sqrt{t}e^{-ct}e^{-h/t}=\frac{1+2\sqrt{ch}}{2c}\sqrt{\frac{\pi}{c}}e^{-2\sqrt{ch}},
\end{equation}
we get finally
\begin{equation} \label{Expon}
S({\bf k})=\frac{2\pi}{3a^2b^3}\sum_{n,m}\left(1+2\frac{b}{a}\Gamma_{nm}\right)e^{-\frac{2b}{a}\Gamma_{nm}}.
\end{equation}

The result of equation (\ref{Expon}) is the main result here, as we got a new series with terms that decay exponentially, rather than the one we started with of equation (\ref{Oscil}) with oscillating terms. Besides giving an intuitive picture this form is particularly useful for numerical calculations as the coefficients are real and decay fast. We can also easily treat some interesting limits in this form.

As an example at this point on we concentrate on long wave length excitons with small $k$. In this limit we have $ka\ll 1$, and hence we can keep only the first term of the series, the one with $(n,m=0)$, which can be easily justified numerically, then we get $\Gamma_{00}=ka/2$, to obtain
\begin{equation}
S({\bf k})\simeq\frac{2\pi}{3a^2b^3}\left(1+kb\right)e^{-kb}.
\end{equation}
We calculate now the exciton dynamical matrix terms. 

We obtain the diagonal elements
\begin{eqnarray}
D_{xx}({\bf k})&=&\frac{2\pi}{a^2}\frac{k_x^2}{k}e^{-kb}, \nonumber \\
D_{yy}({\bf k})&=&\frac{2\pi}{a^2}\frac{k_y^2}{k}e^{-kb}, \nonumber \\
D_{zz}({\bf k})&=&-\frac{2\pi}{a^2}ke^{-kb},
\end{eqnarray}
and the off-diagonal elements
\begin{eqnarray}
D_{xy}({\bf k})&=&D_{yx}^{\ast}({\bf k})=\frac{2\pi}{a^2}\frac{k_xk_y}{k}e^{-kb}, \nonumber \\
D_{xz}({\bf k})&=&D_{zx}^{\ast}({\bf k})=-i\frac{2\pi}{a^2}k_xe^{-kb}, \nonumber \\
D_{yz}({\bf k})&=&D_{zy}^{\ast}({\bf k})=-i\frac{2\pi}{a^2}k_ye^{-kb},
\end{eqnarray}
where $k=\sqrt{k_x^2+k_y^2}$.

Finally we get
\begin{equation}
J^{\prime}({\bf k})=\frac{1}{4\pi\epsilon_0}\frac{2\pi}{a^3}\left\{\left(\mbox{\boldmath$\mu$}_{\|}\cdot\hat{\bf k}\right)^2-\mu_z^2\right\}ka\ e^{-kb}.
\end{equation}
where we defined $\mbox{\boldmath$\mu$}=(\mbox{\boldmath$\mu$}_{\|},\mu_z)$ with $\mbox{\boldmath$\mu$}_{\|}=(\mu_x,\mu_y)$, and the unit vector $\hat{\bf k}={\bf k}/k$.

The in-plane calculations for the exciton dynamical matrix elements, which are given in the appendix, yields
\begin{equation}
J({\bf k})=\frac{F}{4\pi\epsilon_0a^3}\left\{2\mu_z^2-\mu_{\parallel}^2\right\},
\end{equation}
where calculating the first terms of the series (\ref{FFF}) gives $F\simeq 9/2$. Note that the calculation for the limit $ka\ll 1$ in considering only the nearest neighbor interactions gives $F=4$.

Here we present the results for typical optical lattice numbers, where the transition energy is $E_A=1\ eV$, the lattice constant is $a=1000\ \AA$, the transition dipole is $\mu=1\ e\AA$, the interlattices distance is $b=10a$, and the wave number is taken to be $ka=10^{-3}$. We define $J_0=\frac{\mu^2}{4\pi\epsilon_0a^3}$, which is $J_0\approx 1.44\times10^{-8}\ eV$. In figures (2-4) we plot the scaled transfer parameter, $J^{\prime}({\bf k})/J_0$, as a function of the angle between the in-plane wave vector and the transition dipole, $\phi$, where $\left(\mbox{\boldmath$\mu$}_{\|}\cdot\hat{\bf k}\right)=\left|\mbox{\boldmath$\mu$}_{\|}\right|\cos\phi$, for different direction of the transition dipole, where $\mbox{\boldmath$\mu$}=\mu(\sin\theta,0,\cos\theta)$. Figure (2.a) is for $\theta=0$, figure (2.b) for $\theta=\pi/6$, figure (3.a) for $\theta=\pi/5$, figure (3.b) for $\theta=\pi/4$, figure (4.a) for $\theta=\pi/3$, and figure (4.b) for $\theta=\pi/2$. The plots show strong anisotropic effect.

\begin{figure}[h!]
\centerline{\epsfxsize=4cm \epsfbox{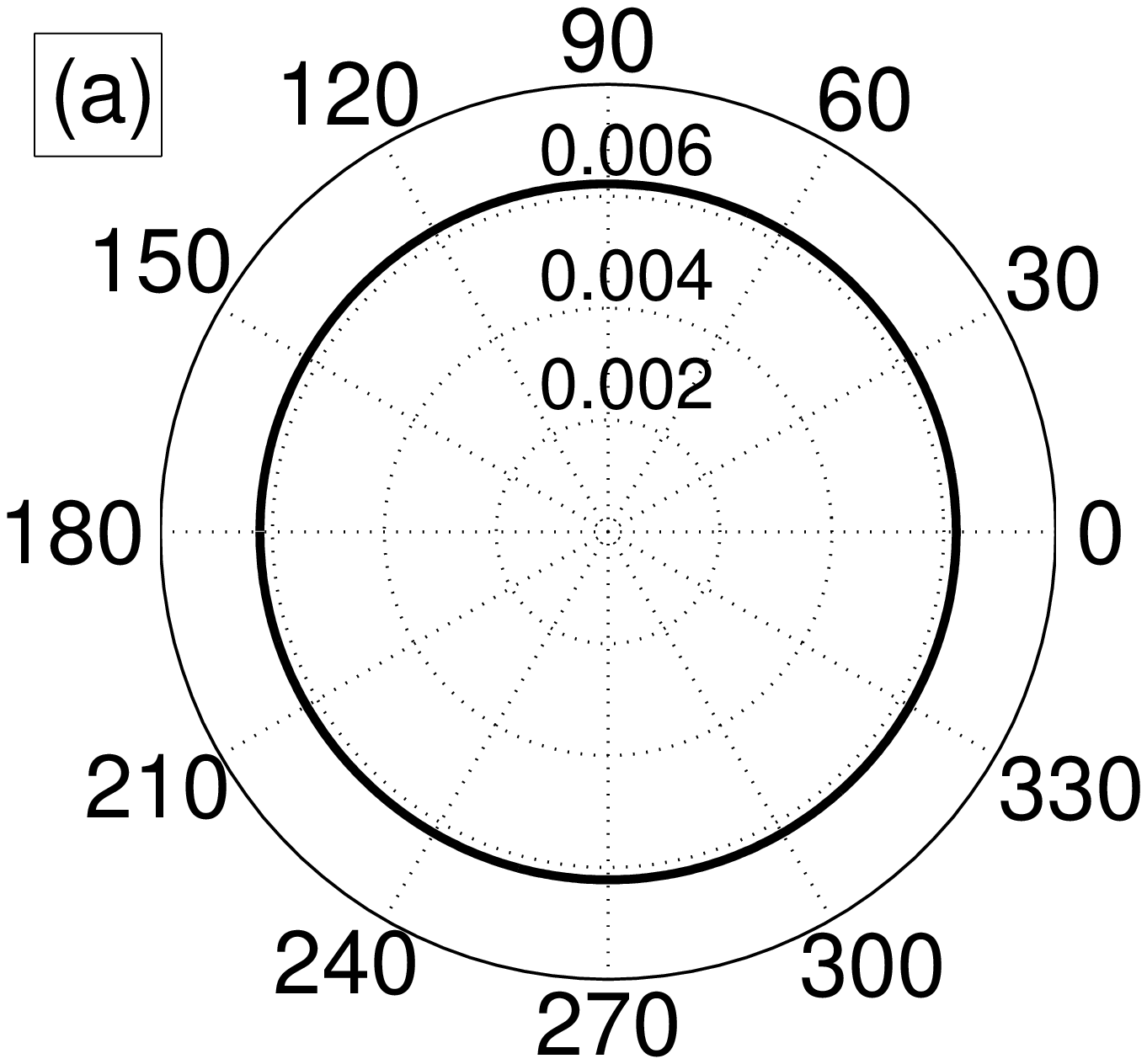}\epsfxsize=4cm \epsfbox{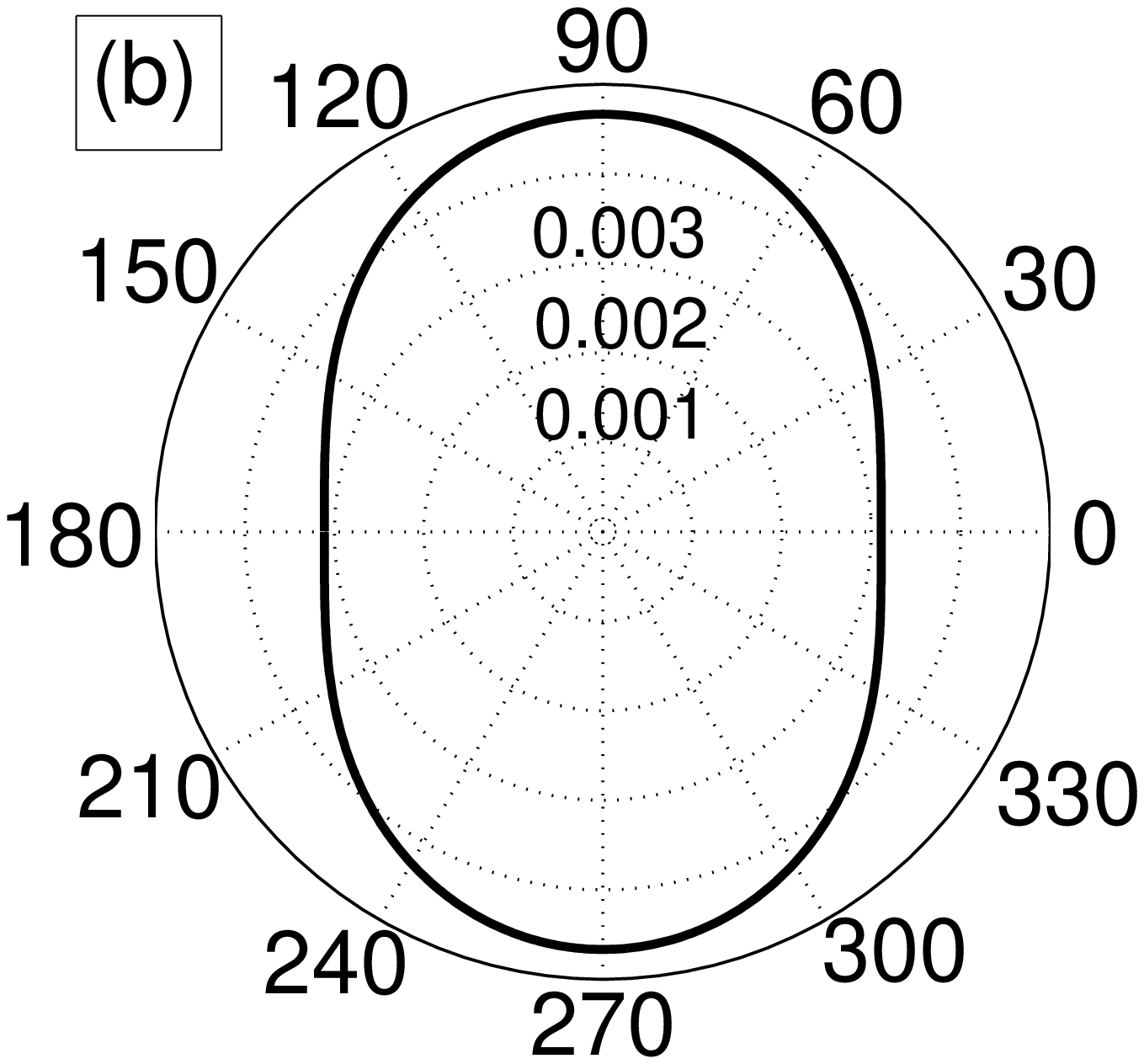}}
\caption{The scaled transfer parameter, $J^{\prime}({\bf k})/J_0$ vs. the angle between the wave vector and the transition dipole, $\phi$, for transition dipole direction of: (a) $\theta=0$, and (b) $\theta=\pi/6$.}
\end{figure}

\begin{figure}[h!]
\centerline{\epsfxsize=4cm \epsfbox{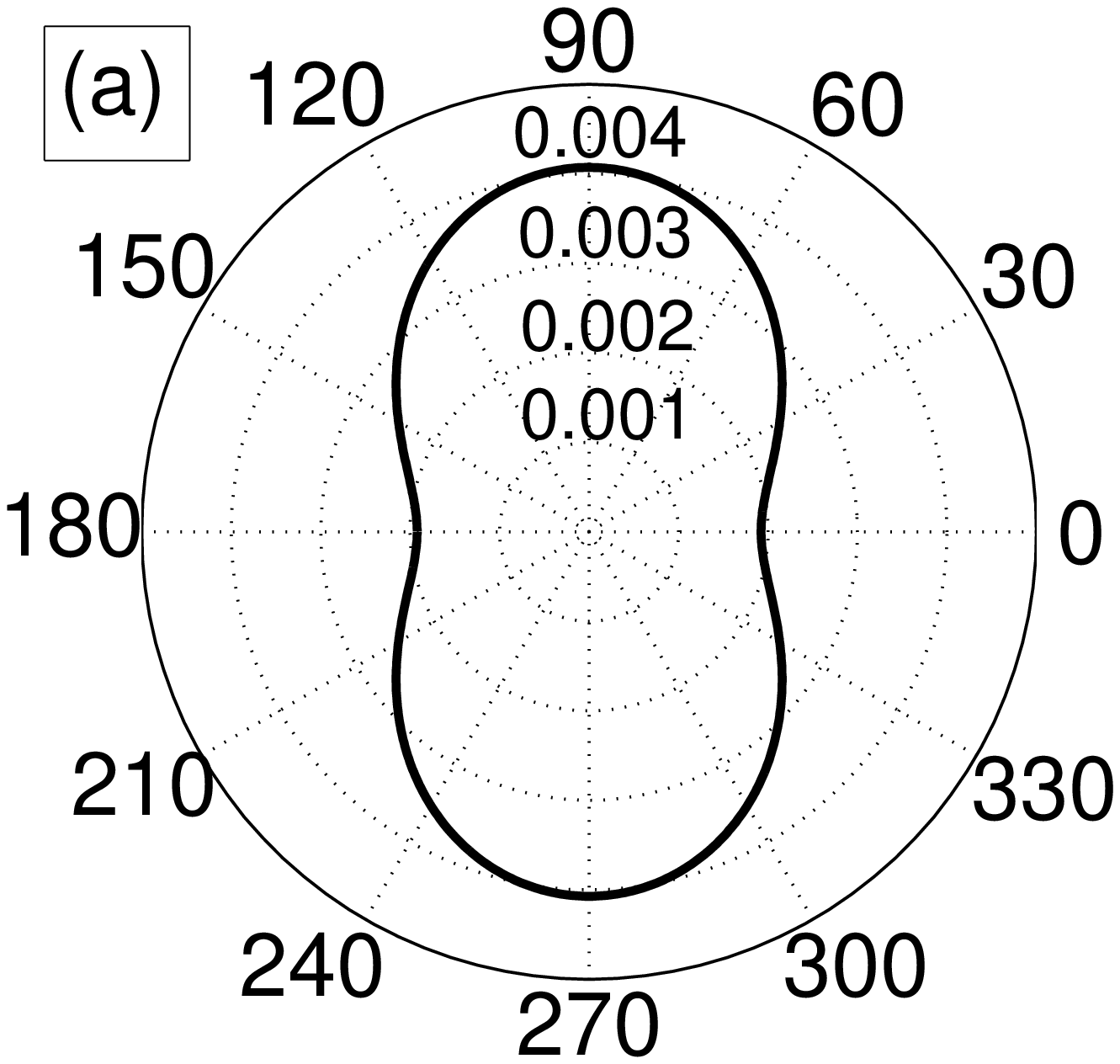}\epsfxsize=4cm \epsfbox{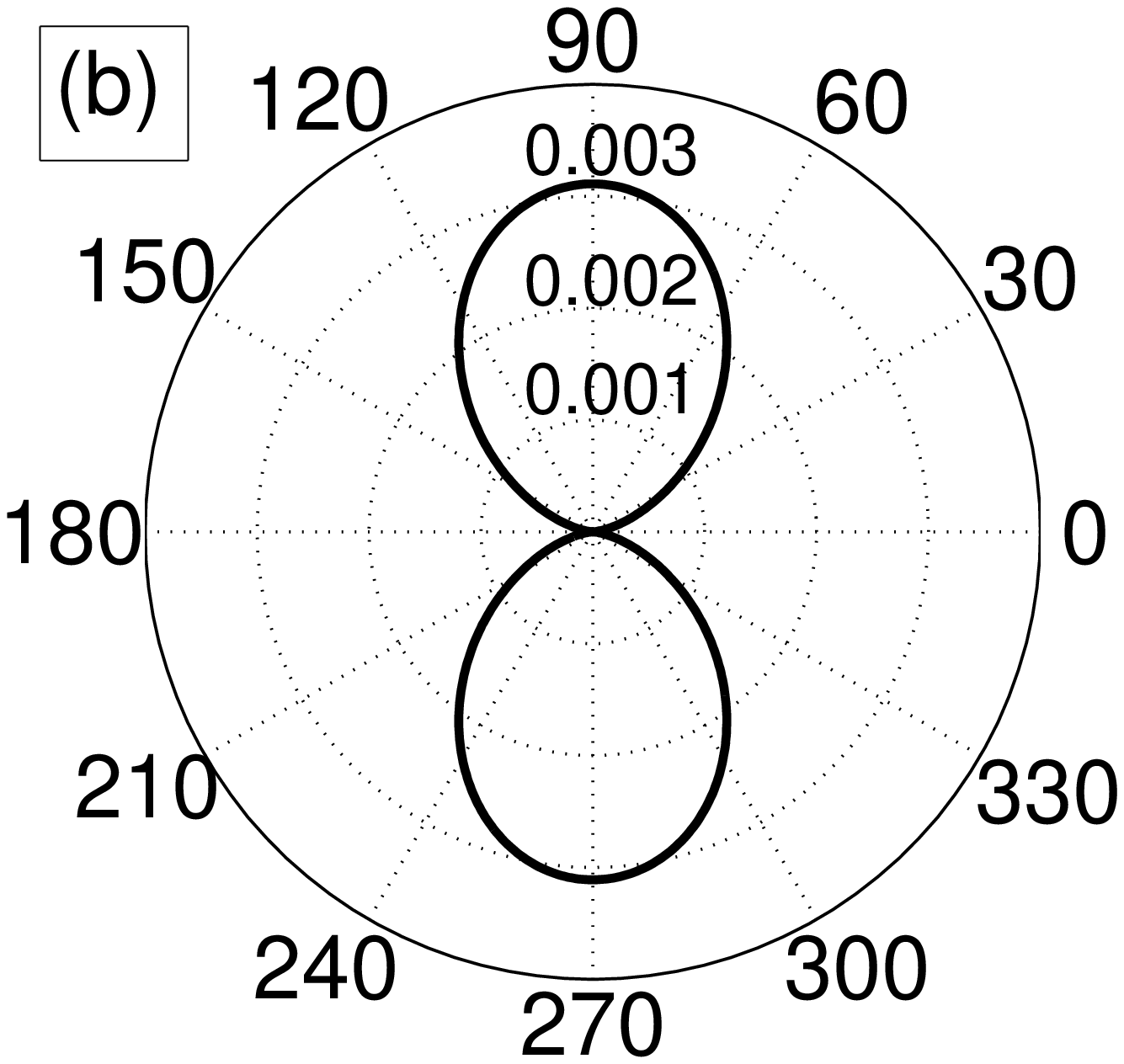}}
\caption{The scaled transfer parameter, $J^{\prime}({\bf k})/J_0$ vs. the angle between the wave vector and the transition dipole, $\phi$, for transition dipole direction of: (a) $\theta=\pi/5$, and (b) $\theta=\pi/4$.}
\end{figure}

\begin{figure}[h!]
\centerline{\epsfxsize=4cm \epsfbox{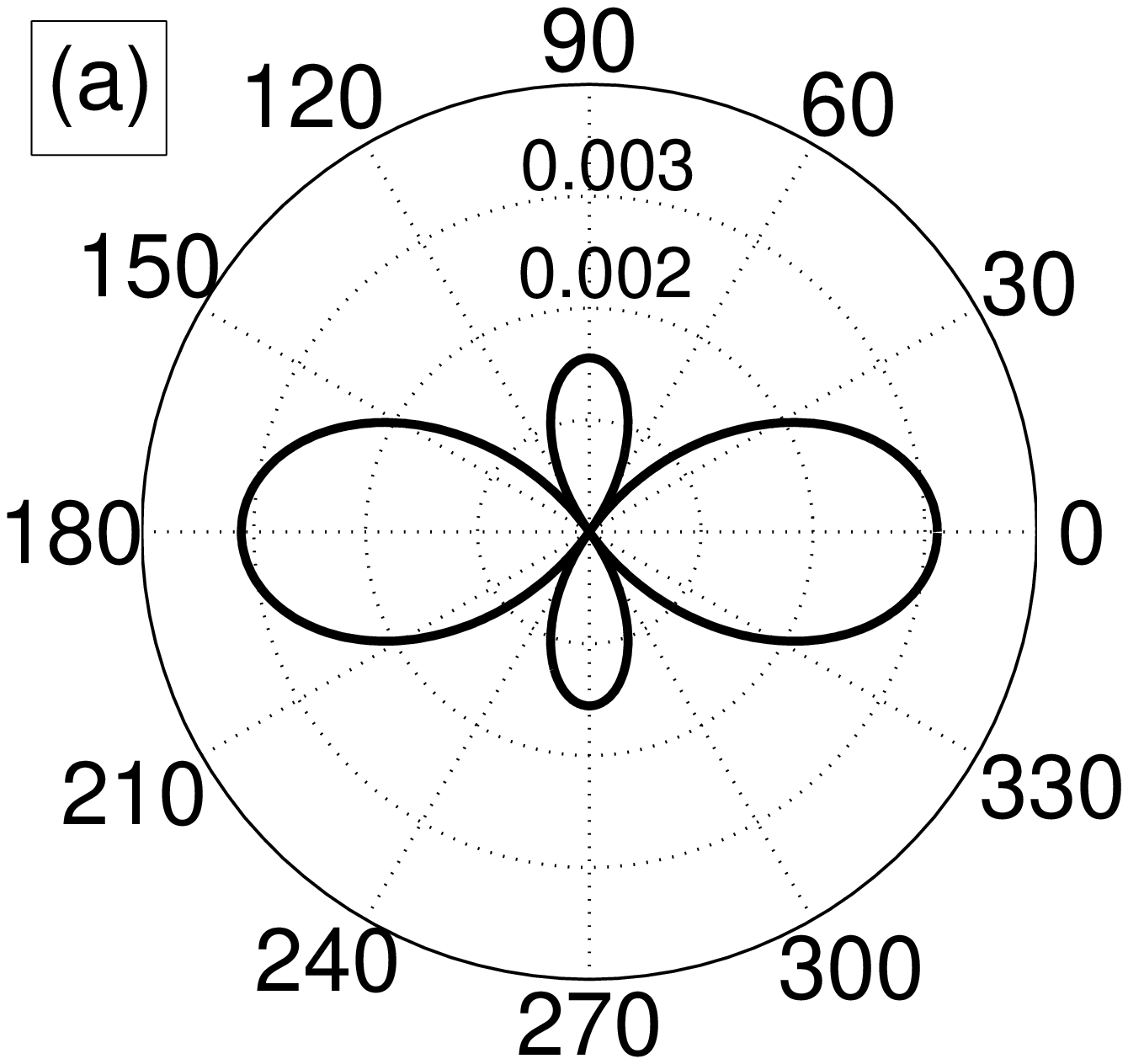}\epsfxsize=4cm \epsfbox{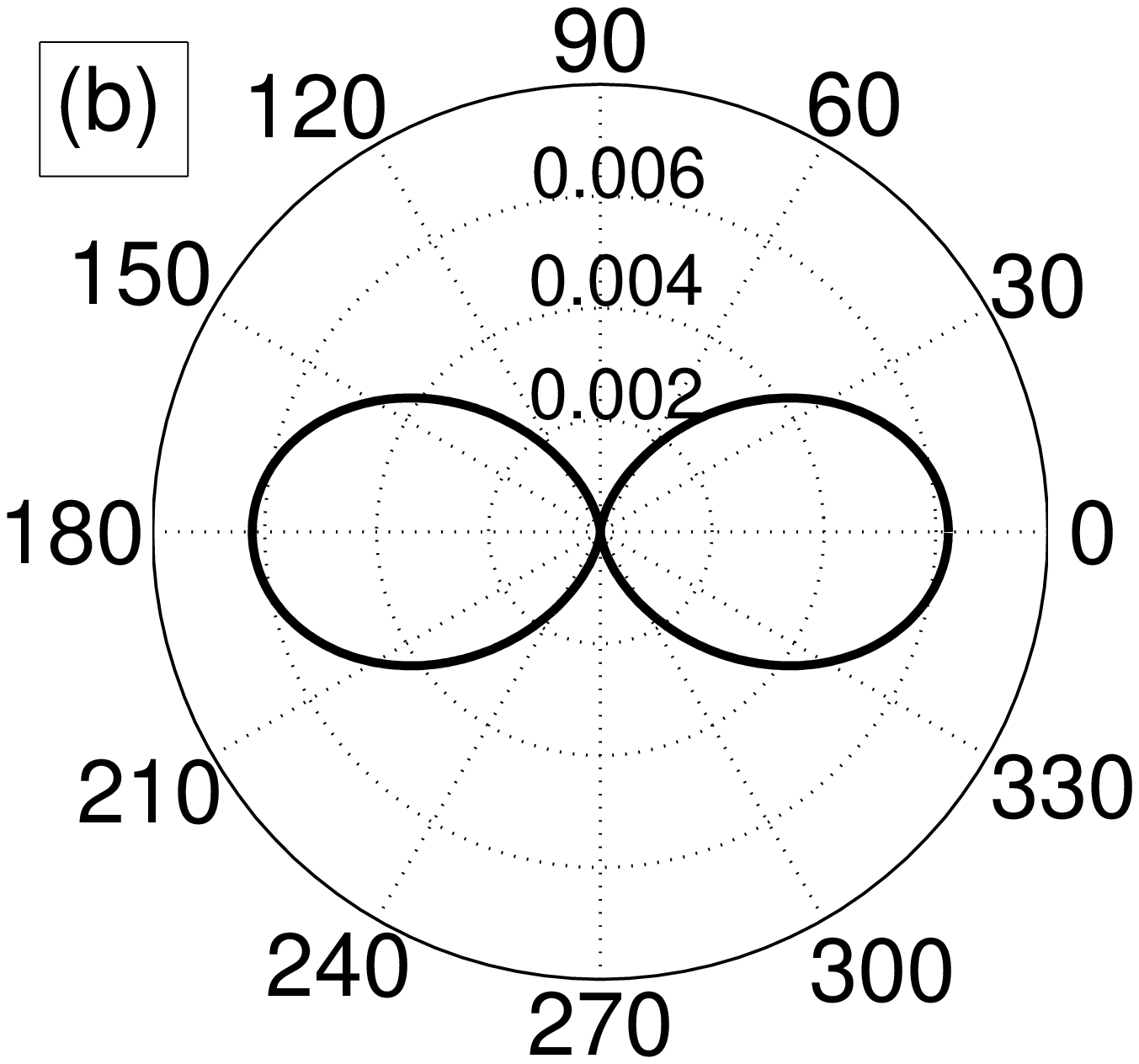}}
\caption{The scaled transfer parameter, $J^{\prime}({\bf k})/J_0$ vs. the angle between the wave vector and the transition dipole, $\phi$, for transition dipole direction of: (a) $\theta=\pi/3$, and (b) $\theta=\pi/2$.}
\end{figure}

\section{Conclusions}

Here we give a discussion of the results we obtained and present some conclusions. For the system of two identical and parallel optical lattices, an exciton that propagates in one lattice with a fixed wave vector ${\bf k}$ can hop to the other one due to dipole-dipole interactions. Such a coupling can form hybrid excitons among the two optical lattices, where we get two hybrid exciton modes, symmetric and antisymmetric modes, which are separated by a splitting energy of $\Delta({\bf k})=E_{ex}^+({\bf k})-E_{ex}^-({\bf k})=2J^{\prime}({\bf k})$, where we found the result
\begin{equation}
\Delta({\bf k})=\frac{1}{\epsilon_0a^3}\left\{\left(\mbox{\boldmath$\mu$}_{\|}\cdot\hat{\bf k}\right)^2-\mu_z^2\right\}ka\ e^{-kb}.
\end{equation}
The result shows strong anisotropic effect in changing the direction of ${\bf k}$ relative to the in-plane transition dipole $\mbox{\boldmath$\mu$}_{\|}$, as seen in figures (2-4). Furthermore, polarization splitting appears at $k=0$ between the normal component $\mu_z$ and the in-plane component $\mbox{\boldmath$\mu$}_{\|}$, which equals $\frac{3F}{4\pi\epsilon_0a^3}$.

The most important feature here is the exponential decay behavior as a function of the inter-lattices distance of the exciton coupling among the two optical lattices. Even though the resonance dipole-dipole interaction has inverse cubic dependence of the distance between two localized dipoles, the appearance of excitons induce exponentially decrease dependence for the exciton collective dipole among the two optical lattices.

For the limit of $b\gg a$ we can neglect the hopping of excitons and it is a good approximation to consider the two lattices independently. While for $b\sim a$ we get the above hybrid symmetric and antisymmetric excitons. Note that the antisymmetric excitons are dark and metastable with long radiative life time, while the symmetric excitons are bright with damping rate two times larger than an exciton in a single optical lattice. Inside a cavity, in the strong coupling regime, the antisymmetric excitons decouple to the cavity photons, while the symmetric excitons are coherently mixed with the cavity photons to form cavity polaritons. We treated similar case in our previous works \cite{ZoubiA,ZoubiB,ZoubiC}.

The present set-up can be generalized to include finite number of parallel planar optical lattices. Then we get hybrid exciton modes as the number of the optical lattices, where part of the modes are dark and others bright. The above important result can be adopted in order to simplify the treatment by assuming hopping of excitons only among nearest neighbor optical lattices. Note that the hopping can be controlled by changing the distance between the optical lattices. The results hold for any similar system of two or multi layers of optically active lattices, e.g., for monolayers of molecular crystals, or lattices of quantum dotes.

The results of the present paper can be adopted, with the appropriate modifications, for dipolar bosonic ultracold quantum gases in optical lattices, which interact via dipole-dipole interactions \cite{Baranov,Pfau}. One candidate for such a set-up is heteronuclear polar molecules that have permanent dipoles. The molecules need to be in their lowest rovibrational states, and an external electric field is applied to orient the molecules along a fixed direction. The other candidate is atomic species with a large magnetic moment that experience magnetic dipole-dipole- interactions, e.g., a BEC of $^{52}$Cr, which have been realized recently in low dimensional optical lattices \cite{Santos}.

\ 

The work was supported by the Austrian Science Funds (FWF), via the project (P21101).

\appendix

\section{Intra-Lattice Exciton Dynamical Matrix}

Here we present the detail calculation for the exciton dynamical matrix elements for the in-plane interactions in using the long range resonance dipole-dipole interactions and in applying Ewald's method \cite{Born}.

Now we take $l_z=0$ in equation (\ref{DDI}). We define
\begin{equation}
S_j({\bf k})=\sum_{l_x,l_y}^{\ \ \ \ \prime}\frac{l_j^2}{\left(l_x^2+l_y^2\right)^{5/2}}e^{ia(k_xl_x+k_yl_y)},\ (j=x,y),
\end{equation}
where the prime on the summation excludes the term with $l_j=0$. The exciton diagonal dynamical matrix elements are
\begin{eqnarray}
D_{xx}({\bf k})&=&-2S_x({\bf k})+S_y({\bf k}), \nonumber \\
D_{yy}({\bf k})&=&-2S_y({\bf k})+S_x({\bf k}), \nonumber \\
D_{zz}({\bf k})&=&S_x({\bf k})+S_y({\bf k}).
\end{eqnarray}
In the following we concentrate in the calculation of $S_x({\bf k})$, as the one for $S_y({\bf k})$ is similar. Using relation (\ref{Rel1}) yields
\begin{eqnarray}
S_x({\bf k})&=&\frac{4}{3\sqrt{\pi}}\int_0^{\infty}dt\ t^{3/2}\left\{\sum_{l_x}^{\ \ \ \ \prime} l_x^2e^{iak_xl_x-l_x^2t}\right\} \nonumber \\
&\times&\left\{\sum_{l_y}e^{iak_yl_y-l_y^2t}\right\}.
\end{eqnarray}
Applying relation (\ref{Rel2}) yields
\begin{eqnarray}
S_x({\bf k})&=&\frac{8}{3}\sum_{l_x=1}^{+\infty}\sum_{n=-\infty}^{+\infty}\cos(ak_xl_x)\left(-\frac{1}{2}l_x\frac{\partial}{\partial l_x}\right) \nonumber \\
&\times&\left\{\int_0^{\infty}dt\ e^{-l_x^2t}e^{-\frac{1}{t}\left(\pi n+\frac{ak_y}{2}\right)^2}\right\}.
\end{eqnarray}
We use now
\begin{equation} \label{Rel3}
\int_0^{\infty} dt e^{-c/t}e^{-ht}=2\sqrt{\frac{c}{h}}K_1(2\sqrt{ch}),
\end{equation}
where $K_n(x)$ is the modified Bessel function of order $n$. We achieve the expression
\begin{eqnarray}
S_x({\bf k})&=&-\frac{4}{3}\sum_{l_x=1}^{+\infty}\sum_{n=-\infty}^{+\infty}\cos(ak_xl_x)\frac{\Lambda_{nl_x}^{k_y}}{l_x^2} \nonumber \\
&\times&\left\{\Lambda_{nl_x}^{k_y}K_1^{\prime}(\Lambda_{nl_x}^{k_y})-K_1(\Lambda_{nl_x}^{k_y})\right\},
\end{eqnarray}
where the prime denotes derivative relative to $l_x$, and
\begin{equation}
\Lambda_{nl_x}^{k_y}=2l_x\left(\pi n+\frac{ak_y}{2}\right).
\end{equation}
We use the relations
\begin{eqnarray} \label{Besel}
K_n^{\prime}(x)&=&-\frac{1}{2}\left\{K_{n-1}(x)+K_{n+1}(x)\right\}, \nonumber \\K_{n+1}(x)&=&K_{n-1}(x)+\frac{2n}{x}K_{n}(x),
\end{eqnarray}
to get
\begin{eqnarray}
S_x({\bf k})&=&\frac{8}{3}\sum_{l_x=1}^{+\infty}\sum_{n=-\infty}^{+\infty}\cos(ak_xl_x)\frac{\Lambda_{nl_x}^{k_y}}{l_x^2} \nonumber \\
&\times&\left\{\frac{\Lambda_{nl_x}^{k_y}}{2}K_0(\Lambda_{nl_x}^{k_y})+K_1(\Lambda_{nl_x}^{k_y})\right\}.
\end{eqnarray}

We move to calculate the off-diagonal term. We define
\begin{equation}
D_{xy}({\bf k})=\frac{3}{a^2}\frac{\partial^2}{\partial k_x\partial k_y}\sum_{l_x,l_y}^{\ \ \ \ \prime}\frac{e^{ia(k_xl_x+k_yl_y)}}{\left(l_x^2+l_y^2\right)^{5/2}}.
\end{equation}
Using the relation (\ref{Rel1}) we get
\begin{eqnarray}
D_{xy}({\bf k})&=&\frac{4}{a^2\sqrt{\pi}}\frac{\partial^2}{\partial k_x\partial k_y}\int_0^{\infty}dt\ t^{3/2}\left\{\sum_{l_x}^{\ \ \ \ \prime} e^{iak_xl_x-l_x^2t}\right\} \nonumber \\
&\times&\left\{\sum_{l_y}e^{iak_yl_y-l_y^2t}\right\},
\end{eqnarray}
and relation (\ref{Rel2}) gives
\begin{eqnarray}
D_{xy}({\bf k})&=&8\sum_{l_x=1}^{+\infty}\sum_{n=-\infty}^{+\infty}l_x\left(\pi n+\frac{ak_y}{2}\right)\sin(al_xk_x) \nonumber \\
&\times&\int_0^{\infty}dt\ e^{-l_x^2t}e^{-\frac{1}{t}\left(\pi n+\frac{ak_y}{2}\right)^2},
\end{eqnarray}
and in using (\ref{Rel3}) we obtain
\begin{equation}
D_{xy}({\bf k})=4\sum_{l_x=1}^{+\infty}\sum_{n=-\infty}^{+\infty}\frac{\Lambda_{nl_x}^{k_y2}}{l_x^2}\sin(al_xk_x)K_1(\Lambda_{nl_x}^{k_y}).
\end{equation}

Here we going to calculate the result for the limit of long wavelength excitons, namely for $ka\ll1$. We start from the recent result of the off-diagonal term, in this limit we keep only the $n=0$ term, to get $D_{x,y}(ka\ll 1)\rightarrow 0$. Note that for $x\gg 0$ we have $K_n(x\rightarrow +\infty)\sim \sqrt{\frac{\pi}{2x}}e^{-x}$ and the series converges.

We return now to the diagonal terms. In the limit of long wavelength excitons we consider the two main contributions:

(i). For $n=0$, in using $\cos(ak_xl_x)\rightarrow 1$, we get
\begin{equation}
S_x(n=0)=\frac{4}{3}\sum_{l_x=1}^{+\infty}k_y^2a^2\left\{K_0(ak_yl_x)+\frac{2}{ak_yl_x}K_1(ak_yl_x)\right\},
\end{equation}
and using (\ref{Besel}) yields
\begin{equation}
S_x(n=0)=\frac{4}{3}\sum_{l_x=1}^{+\infty}k_y^2a^2K_2(ak_yl_x).
\end{equation}
We use also the relation
\begin{equation}
K_n(x\ll 1)\sim \frac{\Gamma(n)}{2}\left(\frac{2}{x}\right)^n,
\end{equation}
where $\Gamma(n)$ is the Gamma function that takes $\Gamma(2)=1$, to obtain the form
\begin{equation}
S_x(n=0)=\frac{8}{3}\sum_{l_x=1}^{+\infty}\frac{1}{l_x^2}=\frac{8}{3}\frac{\pi^2}{6}=\frac{4\pi^2}{9}.
\end{equation}

(ii). For $k=0$ we get
\begin{eqnarray}
S_x(k=0)&=&\frac{16}{3}\sum_{l_x=1}^{+\infty}\sum_{n=-\infty}^{+\infty\ \prime}(\pi n)^2 \nonumber \\
&\times&\left\{K_0(2\pi nl_x)+\frac{2}{2\pi nl_x}K_1(2\pi nl_x)\right\},
\end{eqnarray}
where the prime exclude $(n=0)$. In using (\ref{Besel}) we have
\begin{equation}
S_x(k=0)=\frac{32\pi^2}{3}\sum_{n,m=1}^{+\infty}n^2K_2(2\pi nm).
\end{equation}

Finally we get the result
\begin{equation}
S_x(ka\ll 1)=S_y(ka\ll 1)\simeq F,
\end{equation}
where
\begin{equation}\label{FFF}
F=\frac{4\pi^2}{9}+\frac{32\pi^2}{3}\sum_{n,m=1}^{+\infty}n^2K_2(2\pi nm).
\end{equation}

For the diagonal dynamical matrix elements of excitons we obtain finally
\begin{equation}
D_{xx}=D_{yy}=-F,\ D_{zz}=2F.
\end{equation}

\end{document}